\documentclass[12pt]{article} % use larger type; default would be 10pt

\usepackage[utf8]{inputenc} % set input encoding (not needed with XeLaTeX)
\usepackage{geometry} % to change the page dimensions
\usepackage{graphicx} % support the \includegraphics command and options

\usepackage{physics}
\usepackage{tikz}
\usetikzlibrary{quantikz}
\usepackage{bm}
\usepackage{amssymb}
\usepackage{authblk}
\usepackage{hyperref}
\hypersetup{pdftex,colorlinks=true,allcolors=blue}
\usepackage{hypcap}
\usepackage{algorithm2e}
\usepackage{pgfplots}
\providecommand{\keywords}[1]{\textbf{\textit{Keywords:}} #1}

\title{\textbf{Entaglement-Based Quantum Mean Estimator Circuit}}
\author{Amanuel Tamirat}
\affil{Department of Information Technology, Wolkite University, Ethiopia, amanuel.tamirat@wku.edu.et}

\begin{document}
\maketitle
\begin{abstract}
This paper proposes a quantum circuit for computing the mean value from a given set of quantum states. The circuit consults a Quantum Random Access Memory to get the values of the set, and by using superposition, interference and entanglement phenomena, it can estimate the mean value in  $\mathcal{O}(\frac{1}{\epsilon}\log{Nd})$ complexity. The proposed quantum mean-estimator circuit has been simulated on the IBM Q Experience and the results suggest that the proposed quantum circuit can have the potential to enhance many mean-based machine learning algorithms.
\end{abstract}

\section{Introduction}

Quantum Machine Learning (QML) is an emerging field of a research area and has demonstrated its potential to speed up some of the costly machine learning calculations as compared to the existing classical approaches. For instance, Swap-Test \cite{Esma2006} provides an exponential speedup in the dimension of vectors by calculating distances between two normalized vectors. Following that, in \cite{Lloyd2013} the quantum speedup in the number of vectors during the estimation of euclidian distance from the given vector to the mean of the set operation is shown. Moreover, as described by \cite{Esma2013}, a quadratic speed up by applying quantum minimization algorithm on any distance-based machine learning can be possible \cite{Kopczyk2018}.

Recently, Schuld and her collegues \cite{Schuld2017} proposed a quantum nearest-centroid classifier algorithm based on quantum interference. Given a training dataset, $D = \{(\,x^{1} , y^{1} )\, ,..., (\,x^{M} , y^{M} )\, \}$ of inputs $x^{m}\in \mathbb{R}^{N} $ together with their respective target classes $y^{m} \in \{ 1,-1\}$, the supervised binary pattern classifier can assign the unseen input $\tilde{x}$ to its corresponding class $\tilde{y}$. The algorithm harvests the same logarithmic scaling in the dimension and number of the input data and provides exponential speed up that has been claimed by other authors \cite{Lloyd2013} \cite{Rebentrost2014} \cite{Wiebe2014}.

Many distance-based machine learning algorithms require the calculation of mean from the feature set. Therefore, accelerating such operation for the big data analytics through quantum effect is vital. This paper proposes a quantum circuit for computing the mean value from a given set of vectors $V$. The circuit uses Quantum Random Access Memory (QRAM) to access the values of the vectors parallelly and by using interference and entanglement, it estimates the mean of the set. In the next section, the algorithm that estimates the mean from the QRAM is explained. Then, using the IBM Q Experience the results and the performance of the quantum mean estimator circuit is presented.

\section{Mean Estimator Quantum Circuit}
\begin{figure}[]
\begin{center}
\begin{quantikz}
\lstick{\textit{index register} $\ket{0}^{\otimes n}$} &[2mm] \gate{H}\qwbundle{n} & \gate[wires=2]{Q} & \gate{H} & \octrl{1} & \gate{H}& \qw \\
\lstick{\textit{data register} $\ket{0}^{\otimes m-1}$} & \qw  &  & \qw & \ctrl{1}& \qw & \qw \\
\lstick{\textit{$d_{m}$} $\ket{0}$} & \qw  & \gate{X} & \qw & \ctrl{1}& \qw & \qw &  \\
\lstick{\textit{mean register} $\ket{0}^{\otimes m}$} & \qw  & \qw & \qw & \targ{}& \qw & \qw & \meter{} \\
\end{quantikz}
\end{center}
\caption{Quantum circuit for mean estimation}
\label{fig:mean}
\end{figure}
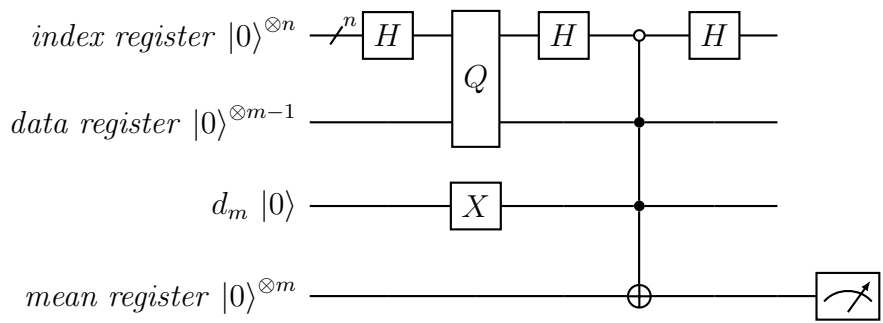

In this section, the proposed quantum mean estimator circuit is described. The general quantum circuit for the mean estimator algorithm is shown in Fig \ref{fig:mean}. Given the size of $N$, $\mathbb{R}^{d}$ quantum states $V$. The proposed quantum circuit returns the average of the quantum states using the interference pattern. As mentioned above, the algorithm requires QRAM to store the values of the given set $V$. QRAM is a device that can (theoretically) encode $N$ d-dimensional classical vectors into (the amplitudes of) a quantum state of $\log{Nd}$ qubits, in time $\mathcal{O}(\log{Nd})$. Hence, it allows us to manipulate them through different quantum gates \cite{Giovannetti2008}. This data representation technique is known as amplitude encoding and it is responsible for most claims of exponential speedups in quantum machine learning algorithms. The QRAM can be simulated as Blackbox by the controlled-rotation around the y-axis by the specified angles $\theta$.

The circuit has three quantum registers namely index register, data register and mean register. The first register contains $n$ qubits where $n = \log{N}$  and it used to represent every $N$ locations in the set. The second and third quantum registers have the same amount of qubits $m$, where $m = \log{d} + 1$. The second quantum register will be used to hold the value of the individual elements of the set, while the third quantum register will be used to hold the mean value.

The algorithm starts by setting up all of the qubits into the ground state, thus, the system state at a time of zero can be exprssed as, $
\ket{\psi_{0}} = \ket{0}^{\otimes n}\ket{0}^{\otimes m}\ket{0}^{\otimes m}
$. At the first step, the Hadamard gate is applied on the index register in order to access the QRAM in superposition state and it puts the system state to $
\ket{\psi_{1}} = \frac{1}{\sqrt{N}}\sum_{i=0}^{N-1}{\ket{i}} \ket{0}\ket{0}
$
Next, the algotihm access the QRAM this makes the system state to include each elements of the set, $\ket{\psi_{2}} = \frac{1}{\sqrt{N}}\sum_{i=0}^{N-1}{\ket{i}} \ket{v_{i}}\ket{0}$. On the third step, the circuit gets the Hadamard gate on the index register only, this brings the state $\ket{\psi}$ to $\ket{\psi_{3}} = \frac{1}{\sqrt{N}}\sum_{i=0}^{N-1} {\frac{1}{\sqrt{N-1}} \sum_{j=0}^{N-1}{(-1)^{i\cdot j}{\ket{j}}} \ket{v_{i}}}\ket{0}$, which is the same as 
$$=\frac{1}{N}\sum_{i=0}{\ket{i}}\sum_{j=0}{(-1)^{i \cdot j}\ket{v_{j}}}\ket{0}$$ 
$$= \frac{1}{N}\left(\ket{0} \sum_{j=0}{\ket{v_{j}}} + \sum_{i=1}{\ket{i}}\sum_{j=0}{(-1)^{i \cdot j}\ket{v_{j}}}\right)\ket{0}$$ Next, if the state of the first register qubits is all zero, the algorithm creates entanglement between the second and the third registers by using a general CNOT gate. The state becomes $$\ket{\psi_{4}}= \frac{1}{N}\left(\ket{0} \sum_{j=0}^{N-1} \sum_{k=0}^{d-1}{v_{jk}\ket{k}\ket{k}} + \sum_{i=1}{\ket{i}}\sum_{j=0}{(-1)^{i \cdot j}\ket{v_{j}}}\ket{0}\right)$$ At last, there is a Hadamard gate that applied on the first register qubits alone, and as a result the system state becomes to 
$$\ket{\psi_{5}}= \frac{1}{N\sqrt{N}} \sum_{i=0} {\ket{i}} \left( \left(   v_{i0}\ket{0} + \sum_{j=1}{v_{ij}-\alpha_{j}\ket{j}}\right)\ket{0} + \sum_{j=1}\alpha_{j}\ket{j}\ket{j}\right)$$ where, $\alpha_{x}= \frac{1}{N}\sum_{i=0}v_{ix}$, and as a result, measuring the third register on the components starting from $d$ upto $2d-1$ yeilds $\abs{\alpha_{j-d}}^{2}$. In case if we want to know the sign of the mean elements, we can rerun the algorithm again by slightly changing or modifying some of the input elements, and measure the results. This allows inspecting which elements were negative and or positive.

\section{Results}

Using the IBM Quantum Experience, two experiments were performed. Both experiments were simulated in a closed-system environment (IBM Quantum Simulator) for the 8192 runs. The underlined quantum algorithm implementation is written in python 3.7 with the help of a module known as QISKit. For a successful implementation of the proposed quantum algorithm, we have implemented a maximum of up to five Controlled-U operations by using a method taken from \cite{Barenco1995}. In the experiments, we have considered only the normalized vectors. Which is the quantum state of the corresponding data vector for $\overrightarrow{u}$  in $V$ by using the equation $\ket{u} = \frac{\overrightarrow{u}}{\abs{u}}$.

\begin{figure}[]
\begin{center}
\begin{tikzpicture}
  \begin{axis}[
    ybar,
    enlargelimits=0.15,
    legend style={at={(0.5,-0.2)},
      anchor=north,legend columns=-1},
    ylabel={Probabilities},
    xlabel={States},
    symbolic x coords={00,01,10,11},
    xtick=data,nodes near coords, 
    nodes near coords align={vertical},
    x tick label style={rotate=45,anchor=east},]
    \addplot coordinates {(00,0.371459961) (01,0) (10,0.336791992) (11,0.291748047)};
  \end{axis}
\end{tikzpicture}
\end{center}
\caption{Result of the IBM Quantum Simulator}
\label{fig:result1}
\end{figure}
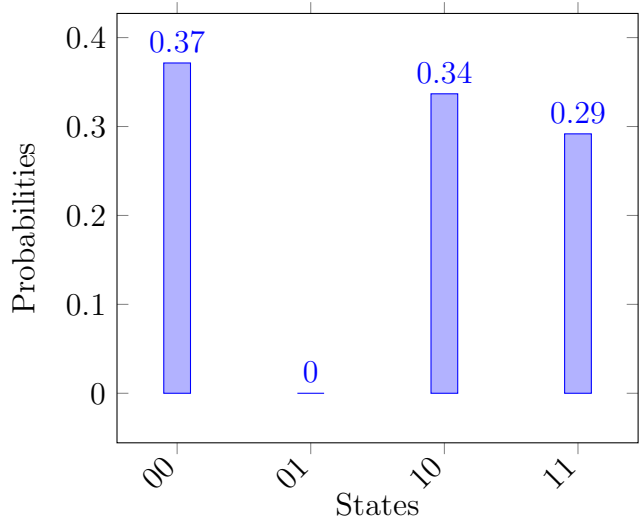

The input variables for the first experiment are listed in Table \ref{table:exp1}, and the result of the simulation is shown in Figure \ref{fig:result1}. As shown in the figure, if we put the probability of the mean register being in 10 and 11 states in square root, we get the mean of the given set with a little error overhead, which is $\sqrt{0.336} = 0.5803$ and $\sqrt{0.291} = 0.5401$.

\begin{table}[]
\caption{Experimental setup of eight two-dimensional quantum states}
\begin{center}
\begin{tabular}{lllll}
\hline
\hline
 \multicolumn{2}{c} {Input vectors}   \\
 \multicolumn{1}{c}{$\theta$ (in rad)} &  \multicolumn{1}{c}{Normalized vectors}   \\
 \hline
 1.5707963267948968 & [0.7071067811865475, 0.7071067811865476]    \\
 0.9272952180016123 & [0.8944271909999159, 0.447213595499958]   \\
 2.3005239830218627 & [0.4082482904638632, 0.9128709291752768]    \\
 1.0471975511965979 & [0.8660254037844386, 0.5]   \\
 1.965587446494658 & [0.5547001962252291, 0.8320502943378437]   \\
 -1.0808390005411688 & [0.8574929257125441, -0.5144957554275267]   \\
 1.21108932720994 & [0.8221921916437787, 0.5692099788303082]   \\
 4.0849933355795707 & [-0.5040655448905066, 0.8636654019087697]  \\
 \hline 
\multicolumn{2}{c} {Classical mean: [0.5819739224925823, 0.5430940952773741]}\\
\hline
\hline
\end{tabular}
\end{center}

\label{table:exp1}
\end{table}

\begin{figure}[]
\begin{center}
\begin{tikzpicture}
  \begin{axis}[
    ybar,
    enlargelimits=0.15,
    legend style={at={(0.5,-0.2)},
      anchor=north,legend columns=-1},
    ylabel={Probabilities},
    xlabel={States},
    symbolic x coords={000, 100,101,110,111},
    xtick=data,nodes near coords, 
    nodes near coords align={vertical},
    x tick label style={rotate=45,anchor=east},]
    \addplot coordinates {(000,0.4156494140625) (100,0.283203125) (101,0.1507568359375) (110,0.08837890625) (111,0.06201171875)};
  \end{axis}
\end{tikzpicture}
\end{center}
\caption{Result of Experiment 2 on IBM Q simulator}
\label{fig:result2}
\end{figure}
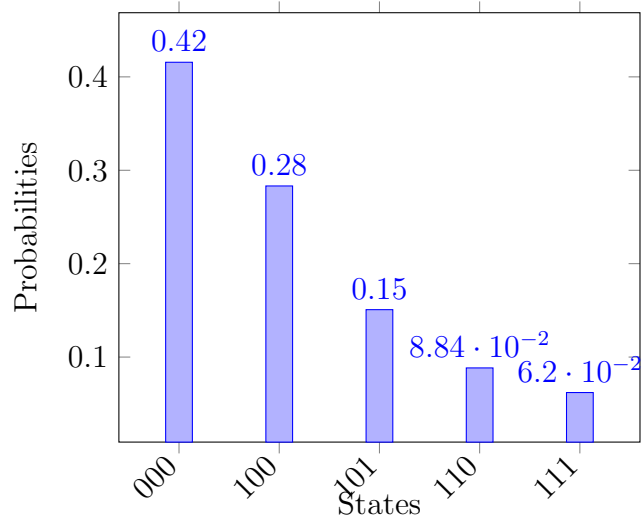

In the second experiment, we have increased the size of the set up to 16 and the dimension of the vectors up to 4. Table \ref{table:exp2} shows the input variables for this experiment. In this case, the algorithm required a total of 10 qubits, where 4 of them found in the index register and 3 qubits each for the data and mean registers were used. The probability of measuring the mean register is shown in Figure \ref{fig:result2}, again if we extract the probability measure of 100, 101, 110 and 111 states, and put them in to square root we can estimate the mean of the set precisely.

\begin{table}[h!]
\caption{Experimental setup of 16 four-dimensional quantum states}
\begin{center}
\begin{tabular}{lllll}
\hline
\hline
 \multicolumn{2}{c} {Input vectors}   \\
 \multicolumn{1}{c}{$\theta_{0}$ and $\theta_{1}$ (in rad)} &  \multicolumn{1}{c}{Normalized vectors}   \\
 \hline
1.965587446494658, $\frac{\pi}{2}$ & [0.3922, 0.3922, 0.5883, 0.5883]    \\
1.0808390005411688, 0.927295218001612 & [0.7669, 0.3834, 0.4601, 0.2300]   \\
$\frac{\pi}{2}$, $\pi$ & [0, $\frac{\sqrt{2}}{2}$, 0, $\frac{\sqrt{2}}{2}$]    
\\
1.21108932720994, 2.3005239830218627 & [0.3356, 0.7505, 0.2323, 0.5196]  \\
2.0849933355795707, -1.0471975511965979 & [0.4365, -0.2520, 0.7479, -0.4318] \\
1.1616780010823367, 4.7243401093344528 & [-0.5946, 0.5876, -0.3903, 0.3856] \\
$\frac{\pi}{2}$, $\frac{\pi}{16}$ & [0.7037, 0.0693, 0.7037, 0.0693] 
\\
1.2293259038443314, 1.6435011087932846 & [0.5563, 0.5982, 0.3926, 0.4223]  \\
1.965587446494658, 1.0808390005411688 & [0.4757, 0.2854, 0.7134, 0.4280]\\
1.21108932720994, 3.0849933355795707 & [0.02326, 0.8218, 0.0161, 0.5689]\\
$\frac{\pi}{8}$, $\frac{\pi}{4}$ & [0.9061, 0.3753, 0.1802, 0.0746]
\\
0.5547001962252291, 0.8320502943378437 & [0.8797, 0.3886, 0.2504, 0.1106]\\
0.4472135954999579, 0.8944271909999159 & [0.8792, 0.4216, 0.1999, 0.0958]\\
0.7808688094430304, 0.6246950475544243 & [0.8799, 0.2841, 0.3621, 0.1169]\\
0.39391929857916763, 0.9191450300180578 & [0.8789, 0.4349, 0.1753, 0.0868]\\
$\frac{\pi}{64}$, $\frac{\pi}{128}$ & [0.9996, 0.0122, 0.0245, 0.0003]
\\
\hline 
\multicolumn{2}{c} {Classical mean: [0.53245, 0.39130, 0.29107, 0.24831]}\\
\hline
\hline
\end{tabular}
\end{center}

\label{table:exp2}
\end{table}

\section{Conclusion}

In this paper, we have used the powerful computing abilities of a quantum computer to find the mean value from the given set of quantum states, the proposed quantum mean estimator circuit runs in a 
$\mathcal{O}(\frac{1}{\epsilon}\log{N2d}+2)$ complexity and as compared to the classical version of it which takes $\mathcal{O}(Nd)$, the proposed algorithm provides an exponential speedup. First, the algorithm access the QRAM in parallel, then by using interference, it calculates the sum of the set. Next, the circuit entangles the sum of the set to the value holder register, this guarantees that upon the measurement of the value holder register, we can estimate the mean value precisely. Due to the normalization property of the quantum states, the returned quantum state is also normalized. For this reason, the proposed algorithm uses two extra qubits to shift $d$ steps towards to the right of the mean value. Alternatively, by padding zero at the first index on every vector in the set, we can avoid this ancilla qubits. This work also shows some experiments on thirty-two qubits remote IBM Quantum simulator. The experiments utilize elementary examples for testing the working performance of the algorithm. The results from the simulation suggest that the quantum mean estimator algorithm can help the central problem of many mean-based algorithms such as k-means, k-medians, and nearest-centroid by calculating the mean value of their feature set quantumly.

\bibliographystyle{unsrt}
\bibliography{myref1}

\end{document}